
\def\doublespace{\baselineskip=20pt plus 2pt\lineskip=3pt minus
     1pt\lineskiplimit=2pt}

\def\nofirstpagenoten{\nopagenumbers\footline={\ifnum\pageno>0\tenrm
\hss\folio\hss\fi}}
\magnification=1200
\font\normfont=cmr12
\font\bigfont=cmr17

\normfont
\pageno=0
\nofirstpagenoten
{\doublespace {\bf\centerline{ Neutrino Masses in Flipped SU(5)
}}} \vskip 1in
{\centerline {\bigfont \bf G.K. Leontaris, and J.D.
Vergados}} \medskip
{\centerline{\sl Physics Department}
\centerline{\sl University of Ioannina}
\centerline{\sl Ioannina GR--451 10}
\centerline{\sl Greece}}
\rm
\vskip 1.5in
\centerline{\underbar {ABSTRACT}}
\smallskip
\par
We analyse the fermion masses and mixings in the flipped SU(5)
model.
The fermion mass matrices are evolved from the GUT scale down
to $m_W$ by solving the renormalization group equations for the
Yukawa couplings. The constraints imposed by the charged
fermion data are then utilised to make predictions about the
neutrino properties . It is found that the {\it
generalized } see-saw mechanism which occurs naturally in this
model can provide {\it i})a solution to the solar neutrino
problem via the MSW mechanism and {\it ii})a sufficiently large
$\nu _{\tau }$ mass to contribute as a hot dark matter
component  as indicated by the recent COBE data.
 \footnote {}
{IOA-290\qquad\qquad\qquad\qquad\qquad\qquad\qquad
\qquad\qquad\qquad\qquad\qquad\qquad December 1992} \vfill
\eject
\pageno=1
\par

Nowadays there is a lot of experimental evidence that
the neutrinos (or at least some of them) might have
non-vanishing--although tiny--masses.  Recent data from
solar neutrino experiments$^{[1]}$  show that the
deficiency of solar neutrino flux, i.e. the discrepancy
between theoretical estimates and the experiment, is
naturally explained if the $\nu _e$ neutrino oscillates to
another species during its flight to the earth. In
addition, new evidence has been reported $^{[2,3]}$
for a significant depletion on the
atmospheric $ \nu_{\mu} $ flux. This can also be explained in
terms of $\nu_{\mu}\longleftrightarrow  \nu_{e}$
 oscillations with mass difference of order
 $ \Delta m^2 \sim 10^{-2} - 10^{-3} eV^2 $ and a relatively large
mixing angle $(sin^2 2\theta _{\tau \mu} \geq .42)$.
Furthermore, the COBE
measurement $ ^{[4]}$ of the large scale
microwave background anisotropy, might be
explained $^{[5]}$ if one assumes an
admixture of COLD ($ \sim 75\% $) plus HOT ($\sim25\% $)
dark-matter. It is hopefully expected that some
 neutrino (most likely $\nu _{\tau }$)
  may be the natural candidate of the hot dark
matter component.

{}From the theoretical point of view, neutrino masses are
zero in the minimal standard model of the electroweak
intetractions. Non-zero neutrino masses arise naturally
however, in most of the Grand Unified Theories (GUTs)
as well as in Supersymmetric ones (SUSY GUTs). In all
 these models, neutrino masses are related to the
quark masses. In general one usually obtains a Dirac
type neutrino mass matrix very similar (or even
identical) to the up-quark mass matrix at the GUT scale.
Small neutrino masses in these models, compatible with
the experimental constraints, are obtained in terms of
the see-saw mechanism$^{[6]}$ . There, the left-handed
 $\nu $, and right-handed
$\nu ^c$ components of the neutrinos form the following mass matrix
$$\left(\matrix{0&m_{\nu _D}\cr
m_{\nu _D}^T&M}\right) \eqno(1)$$
where $m_{\nu _D}$ is the Dirac type 3x3 mass matrix $(\sim m_{up})$
while M is a 3x3 Majorana mass matrix with
entries usually of the order of the GUT scale. After
diagonalization one obtains small left-handed Majorana
 masses of the order of $m_{\nu } \sim m^2_{\nu _D}/M $ and
heavy right-handed majorana states of order $M \sim
M_{GUT}$. Light neutrino eigenmasses may then be evolved
down to low energies and be compared with the experimental
limits. The constraints put by the aforementioned
neutrino data and their relation to the quark masses
at the GUT scale is a real challenge for most of the proposed
GUT models.
Recently, motivated by the observed merging of the Standard
Model gauge coupling constants in SUSY-GUTs there has been a
revived interest in determining the low energy parameters of
the theory in terms of few inputs at the GUT scale$^{[7,8]}$
in the limit of zero neutrino masses. Since however most of the
GUT models  naturally predict the existence of right-handed
neutrinos, the proposed framework has now been expanded
$^{[9,10]}$ to include non-vanishing neutrino masses as well.
The general strategy in these approaches is to use the minimal
number of parameters at the GUT scale so as to have the
maximum number of predictions at $m_W$. Ultimately, one hopes
that this minimal set of parameters at the GUT scale may be
justified in terms of a more fundamental theory, such as the String
Theory. We should point out, however, that not only $m_{\nu _D}$ is
related to the up-quark masses but other indirect constraints come
also from the rest of fermions. It thus appears challenging to
utilize all possible such constraints in the mass matrix of eq.(1)
in order to make definite predictions for the as yet elusive
neutrinos which will then be checked by experiment, this way
supporting or exluding such GUT scenarios.

In the present work we would like to address the question of
fermion and in particular the neutrino masses in GUT models
which arise $^{[11,12]}$in the free fermionic construction of
four dimensional strings. As an application we are going to
consider the $SU(5)\times U(1)$ model of ref[11], but our
results are valid also for the model of ref.[12].
There has been much fruitful work$^{[13,14]}$ in this kind of
models the last few years. Recently it was shown$^{[15]}$ that the
general see-saw mechanism which occurs naturally in this kind of
models, turns out to be consistent with the recent solar neutrino
data, while on the other hand suggests that CHOROUS and NOMAD
experiments at CERN may have a good chance of observing
$\nu_{\mu}\longleftrightarrow  \nu_{\tau }$ oscillations.
Here we are going to explore the neutrino masses in detail, assuming
a specific ansatz for the quark and lepton mass matrices at the GUT
scale, which is more or less dictated by some first attempts in
deriving the above model from the four dimensional free fermionic
Superstrings. As stated above, our general strategy is to use the
minimum number of inputs so as to have the maximum number of
predictions. We are going to make use of the GUT relations in
order to fix these inputs in terms of well known low energy masses
of the charged leptons as well as the $m_u$ and $m_c$ quarks and then
to predict the rest of the fermion mass spectum.

 The various tree-level superpotential mass terms which contribute to
the neutrino mass matrix of the flipped $SU(5)$ model are the
following:
$$\lambda _{ij}^{u}F^{i}\bar f^{j}\bar h +\lambda
_{ij}^{\phi \nu ^{c}}F^{i}\bar H\phi ^{j}+ \lambda _{ij}^{\phi}\phi ^0\phi
^{i}\phi ^{j} \eqno(2)$$
where in the above terms $F^{i},\bar f^{j}$ are the $ {10},\bar 5$
matter SU(5) fields while $\bar H,\bar h, h$ are the $\bar {10},\bar
5, 5$ Higgs representations and $\phi^{i}$ are neutral $SU(5)\times
U(1)$  singlets. The Higgs field $\bar H$ gets a vacuum expectation
value(v.e.v.) of the order of the SU(5) breaking scale ($\sim
10^{16}GeV$), $\bar h, h$ contain the standard higgs doublets while
$\phi ^0$ acquires a v.e.v. most preferably at the electoweak scale.
The neutrino mass matrix may also receive significant contributions
from other sources$^{[16],[13-15]}$. Of crusial importance, are the
non-renormalizable contributions$^{[16]}$ which may give a direct
$M_{\nu ^c\nu ^c}=M^{rad}$ contribution which is absent in the
tree-level potential. Then, the general $9\times 9$ neutrino mass
matrix in the basis $(\nu _i,\nu _i^c,\phi _i)$, may be written as
follows:

$$m_{\nu }=
\left(\matrix{0&m_U&0\cr
m_U&M^{rad}&M_{\nu ^{c},\phi }\cr
0&M_{\nu ^{c},\phi }&\mu _{\phi }\cr}\right) \eqno(3) $$
where it is understood that all entries in eq.(3) represent $3\times
3$ matrices. The above neutrino matrix is different from that of
eq.(1), since now there are three neutral $SU(5)\times U(1)$ singlets
involved, one for each family. Note that this type of neutrino mass
matrix with additional singlets, has also been introduced on purely
phenomenological grounds, in various extensions of the standard
model$^{[17]}$. On the contrary, here the extended see-saw mechanism
is natural in string derived models.

It is clear that the matrix (3) depends on a relatively large number of
parameters and a reliable estimate of the light neutrino masses and the
mixing angles  is a rather complicated task. We are going to use however
 our knowledge of the rest of the fermion spectrum to reduce
sufficiently the number of parameters involved. Firstly, due to the GUT
relation $m_U(M_{GUT})=m_{\nu _D}(M_{GUT})$, we can deduce the form of
$m_{\nu _D}(M_{GUT})$, at the GUT scale in terms of the up-quark masses.
The heavy majorana $3\times 3$ matrix $M^{rad}$, depends on the kind of the
specific generating mechanism. For example, if $M^{rad}$ is due to
some non-renormalizable interactions, then it is completely model
dependent. Here,in order to be specific, motivated by the fact that
in the non-supersymmetric version of the model this matrix may be
generated radiatively$^{[18]}$, we take it to be proportional to the
down quark-matrix $^{[14]}$ at the GUT scale:
$$M^{rad}=\Lambda ^{rad} m_D(M_{GUT})\eqno(4)$$
 The
$M_{\nu ^{c},\phi }$ and $\mu _{\phi }$  $3\times 3$ submatrices are
also model dependent. In most of the string models however, there is
only one entry at the trilinear superpotential in the matrix $M_{\nu
^{c},\phi }$, which is of the order $M_{GUT}$. Other terms, if any,
usually arise from high order non-renormalizable terms. We will
assume in this work  only the existence of the trilinear term,
since higher order ones will be comparable to  $M^{rad}$  and are
not going to change our predictions. In particular we will take
$M_{\nu ^{c},\phi }\sim Diagonal(M,0,0)$, and $\mu _{\phi }\sim
Diagonal(\mu,0,0)$, with $\mu << M\sim M_{GUT}$, thus we will treat
(3) as a $7\times 7$ matrix.

Our ansatz for the other fermion mass matrices is
$$m_U=q
\left(\matrix{0&0&x\cr
0&y&z\cr
x&z&1\cr}\right)\equiv m_{\nu _D} \eqno(6a)$$
$$m_{D}=s \left(\matrix{0&\alpha &0\cr\alpha &b&0\cr
0&0&f\cr}\right), m_{E}= s\left(\matrix{0&\alpha &0\cr
\alpha &-3b&0\cr
0&0&f\cr}\right)\eqno(6b) $$
where
$$s={\upsilon \over\sqrt{2}} sin\beta, q = \lambda _t(t_0) {\upsilon
 \over\sqrt{2}}cos\beta\eqno(6)$$
and $tan\beta ={\bar {\upsilon }\over \upsilon }\equiv {\bar h
\over h}$.
The form of the above mass matrices is considered at the
GUT scale. In order to find their structure at the low energy scale
and calculate the mass eigenstates as well as the mixing matrices and
compare them  with the experimental data, we need to evolve them
down to $m_W$, using the renormalization group equations. Using the
results of Ref.[19] we obtain the renormalization group equations
for the Yukawa couplings at one-loop level
$$16\pi^2{d\over dt} \lambda_U =
(I^{.}  Tr[3\lambda _U \lambda _U^{\dagger}] +
3\lambda _U \lambda _U^\dagger + \lambda _D
\lambda^\dagger_D - I^{.}G_U)\lambda _U ,
\eqno(7)$$

$$16\pi^2{d\over dt} \lambda_N =
(I^{.}Tr[\lambda _U \lambda _U^{\dagger}] +
\lambda _E \lambda _E^\dagger - I^{.}G_N)
\lambda _N , \eqno(8)$$

$$16\pi^2{d\over dt} \lambda_D =
(I^{.}(3Tr[\lambda _D \lambda _D^{\dagger}] +
Tr[\lambda _E \lambda _E^{\dagger}]) + 3\lambda _D
\lambda^\dagger_D + \lambda _U \lambda _U^\dagger - I^{.}
G_D)\lambda _D , \eqno(9)$$

$$16\pi^2{d\over dt} \lambda_E =
(I^{.}(Tr[\lambda _E \lambda _E^{\dagger}] +
Tr[\lambda _D \lambda _D^\dagger])+ 3\lambda _E \lambda _E^\dagger
- I  G_E)\lambda _E , \eqno(10)$$
where $\lambda_\alpha$, $\alpha = U,N,D,E,$ represent the
$3\times3$ Yukawa matrices which are defined in terms of the
mass matrices given in eqs. (4-6), and I is the $3\times3$
identity matrix. We have neglected one-loop corrections
proportional to $\lambda_N^2$. $t \equiv ln(\mu/\mu_0)$, $\mu$
is the scale at which the couplings are to be determined and
$\mu_0$ is the reference scale, in our case the GUT scale.  The
gauge contributions are given by

$$G_\alpha = \sum_ {i=1}^3  c_\alpha^i g_i^2 (t),  \eqno(11)$$

$$ g_i^2 (t) = {g_i^2 (t_0) \over {1- {b_i\over {8\pi^2}}
{g_i^2(t_0)}(t - t_0)}} \eqno(12)$$

The $g_i$ are the three gauge coupling constants of the Standard
Model and $b_i$ are the corresponding beta functions in minimal
supersymmetry.  The coefficients $c_i^\alpha$ are given by

$$\{{c_U^i}\}_{i = 1,2,3} = {\Big\{{13\over{15}}, 3 ,
{16\over3}} \Big\},  \{{c_D^i}\}_{i = 1,2,3} =
\Big\{{7\over{15}}, 3 , {16\over3} \Big\},\eqno(13)$$

$$\{{c_E^i}\}_{i = 1,2,3} =
\Big\{{9\over5}, 3,0 \Big\},
\Big\{{c_N^i}\Big\}_{i = 1,2,3} =
\Big\{{ 3\over5}, 3 ,0 \Big\},\eqno(14)$$
\par
In the following, we find it convenient to redefine the  quark and
lepton fields such that $\lambda_U$ and $\lambda_N$ are
diagonal

$$ \lambda _U  \rightarrow \lambda _U = K^{\dagger}
\lambda _U K, \lambda _N  \rightarrow \lambda _N = K^{
\dagger} \lambda _N K, \eqno(15)$$

The matrix which diagonalizes the up quark mass matrix at the GUT scale
is given by ($x < y < z$)
 $$K=\left(\matrix{{y-z^2\over
D_1}&-{xz\over D_2}&{xy(1-y+z^2)\over{(1+z^2) D_3}}\cr {xz\over
D_1}&{(y-z^2)\over D_2}&{zy\over D_3}\cr
 -{xy\over D_1}&-{z(y-z^2)\over D_2}&{y(1-y+z^2)\over
D_3}\cr}\right)\eqno(16)$$

with
$$\eqalign{D_1\approx \bigl((y-z^2)^2(1+x^2)+x^2z^2\bigr) ^{1\over
2}\cr D_2\approx \bigl((y-z^2)^2(1+z^2)+x^2z^2\bigr) ^{1\over 2}\cr
D_3\approx \bigl(y^2(1-y+z^2)+y^2z^2\bigr) ^{1\over 2}\cr}$$

The mass eigenvalues at the GUT scale read:

$$m_1 \approx q {-x^2y\over y-(x^2+z^2)}, m_2-m_1 \approx
q \left (y - {x^2+z^2\over 1-y} \right), m_3 \approx
q \left(1+ {x^2+z^2\over 1-y} \right)\eqno(17)$$

We apply the field redefinitions (15) to the
differential equations (7-10) and within the
parentheses on the right hand side we retain only the
dominant Yukawa coupling $\lambda _t^2(t)$

$$16\pi^2{d\over dt} \lambda _U = ({\lambda _t^2(t)}
\left(\matrix{3 & { } & { }\cr
{ } & 3 & { }\cr
{ } & { } & 6\cr} \right) - G_U(t)I) \lambda _U  ,\eqno(18)$$

$$16\pi^2{d\over dt} \lambda _N = ( \lambda _t^2( t)
\left(\matrix{1 & { } & { }\cr
{ } & 1 & { }\cr
{ } & { } & 1\cr} \right)
- G_N(t)I)) \lambda _N  ,\eqno(19)$$

$$16\pi^2{d\over dt} \lambda_D = {\lambda _t^2(t)}
\left(\matrix{0 & { } & { }\cr
{ } & 0 & { }\cr
{ } & { } & 1 \cr} \right) - G_D(t)I) \lambda _D  ,\eqno(20)$$

$$16\pi^2{d\over dt} \lambda _E = - G_E(t)I \lambda _E
,\eqno(21)$$

Solving eqs.(18-21), we obtain:

$$\lambda _t(t) = \lambda _t(t_0) {\zeta ^6}
{\gamma _U}(t) \eqno(22)$$

where

$$ {\gamma_{\alpha}(t) = exp(- \int {G_ \alpha} (t) dt/ {(16
\pi^2))}} \eqno(23)$$

$$= \Pi_ {j=1}^3 \left({\alpha _{j,0}\over
\alpha_j}\right)^{c_\alpha ^j/2b_j} \eqno(24)$$

$$= \Pi _{j=1}^3 \left(1- {b_{j,0}\alpha _{j,0}(t-t_0)\over
2\pi} \right)^{c_{\alpha }^j/2b_j} \eqno(25)$$

$$\zeta = exp \left({1\over {16\pi ^2}}  {\int_{t_0}^t}
 \lambda _t^2(t) dt \right) \eqno(26)$$

$$=\left(1- {3 \over 4\pi^2} {\lambda _{\alpha}}{(t_0)}
\int_{t_0}^t \gamma _t^2(t) dt \right)^{-1/12}
\eqno(27)$$

Then, the up quark masses are predicted to be:
$$m_u \approx  \gamma _{_U}  \zeta ^3  q {x^2y \over
y-x^2-z^2}  n_u\eqno(28)$$

$$m_c \approx  \gamma _U  \zeta ^3  q
  \left (y-{z^2+x^2 \over 1-y} \right) n_c+ {\eta _c\over \eta _u}m_u
\eqno(29)$$

$$m_t \approx  \gamma _U  \zeta ^6  q
\left (1 + {x^2+z^2 \over 1-y} \right)\eqno(30) $$

In the above formulae, $\eta _u$ and $\eta _c$ are taking into account
the effects of QCD renormalization from the scale $m_t$ down to 1$GeV$
for $m_u$ and to $m_c$ for $m_c$.

 Similarly, renormalizing ${\lambda _N}$ down to  $m_t$ and
expressing the eigenvalues in terms of the up-quark masses, we find
that the Dirac-neutrino masses are

$$ m_ {{\nu}_{D_1}} \approx  {{\gamma _N} \over {\gamma
_U}} {1 \over {\eta _u} {\zeta ^2}} m_u,m_{{\nu}_{D_2}} \approx  {{\gamma _N}
\over {\gamma
_U}} {1 \over {\eta_ u} {\zeta ^2}} m_c,m_ {{\nu }_{D_3}} \approx  {{\gamma _N}
\over {\gamma
_U}} {1 \over  {\zeta ^5}} m_t \eqno (31)$$

In the above basis where the up quark and neutrino  matrices are diagonal,
the renormalized down quark mass matrix is found to be

$$m_D^{ren} \approx \gamma _D \left(\matrix {1 & { } & { } \cr
{ } & 1 & { } \cr
{ } & { } & {\zeta }} \right) s
 K^{\dagger}\left(\matrix{0 & {\alpha} & 0\cr
{\alpha} & b & 0 \cr
0 & 0 &f} \right) K  \eqno(32)$$

while for the leptons one gets the matrix

$$m_E^{ren} \approx {\gamma _E} s K^{\dagger}
\left(\matrix{0 & {\alpha} & 0\cr
{\alpha} & {-3b} & 0\cr
0 & 0 & f}\right) K\eqno(33)$$

We consider the lepton masses as inputs, and we find the approximate
expressions for the down quarks in terms of the leptons to be:

$$ m_b \approx {{\gamma _D} \over {\gamma _E}}  {\zeta} {m _\tau}
n_b \eqno(34)$$

$$ m_d \approx -n_d   {{\gamma _D} \over
 {6\gamma _E}} \left ( m_{\mu }
 - m_e - \sqrt{(m_{\mu } - m_e)^2
+ 36m_{\mu }m_e} \right )  \eqno(35) $$

$$ m_s \approx n_s    {{\gamma _D} \over {6\gamma
_E}} \left ( m_{\mu }   - m_e + \sqrt{(m_{\mu }
- m_e)^2
+ 36m_{\mu }m_e)} \right ) \eqno(36) $$
where now, $\eta _d$ , $\eta _s$ and $\eta _b$, are taking into
account the QCD renormalization effects for the corresponding down
quarks and $\zeta ,\gamma _{\alpha}$ are given in terms of (23)-(27).
We will take $\eta _{d,s,u}\approx 2, \eta _c \approx 1.8$ and
$\eta _b\approx 1.4$
 Now,
since the range of the charged lepton masses are well known, one can use the
above equations to determine the corresponding range of the down quarks and
compare it with the running masses of $d$,$s$ and $b$. The range  of the
latter, is determined via SU(4) mass relations or QCD sum
rules$^{[20]}$. Thus, for example, from SU(4) mass relations one gets
 $$m_d =7.9\pm 2.4MeV, m_s=155\pm 50MeV,\eqno(37)$$
and from QCD sum rules
 $$m_d = 8.9\pm 2.6MeV, m_s=175\pm 55MeV,\eqno(38)$$
while $m_b = 4.25 \pm .10GeV$. An interesting fact is that  the
renormalization parameter $\zeta $ is constrained in a narrow region
in order to give the correct prediction for the bottom mass. For all the
acceptable $m_t$ range ${\gamma _D \over \gamma _E} \approx 2.1$
and  $\eta _b \approx 1.4 $,
thus  $\zeta \approx .81\pm .2$. The predictions of the other two
down quark masses are $m_s \approx {153 MeV}$ and $m_d \approx {6.3
MeV}$. The $m_s$ value is within the acceptable ranges given in
(37-38). $m_d$ value is somewhat low but still in the range of
(37).

The Kobayashi-Maskawa (KM) matrix can be determined by diagonalizing the
down-quark matrix in (32). In order to determine the KM-mixing angles, we
first determine the values of the parameters of $x,y,z$ which give the
correct masses $ m_u=5.1\pm 1.5MeV, m_c=1.27\pm .05GeV $, while
always we adjust properly $tan\beta $ and $\lambda _t(t_0)$, so as to
obtain the correct value for $m_b$. It is worth noting here that the
restricted region of $\zeta$ has a significant impact on the $m_t$
value. Indeed, as $m_t$ gets smaller, the range $(M_{GUT}-m_t)$
becomes bigger, thus the value of $\zeta $ increases. For
$M_{GUT}\approx 10^{16}GeV$,(using  $\eta _b \approx 1.4 $), we find
that $m_b$ is pushed to its upper limit, when  $m_t$ is around
$125GeV$.$m_b$ goes to its lower limit as $m_t$ approaches
$175GeV$, while $m_u$($ m_c$) gets its lower(higher)acceptable
value.  We also keep track of the ratios $15\le {m_s\over m_d} \le
25,.2\le {m_u\over m_d} \le .7$, which are constrained by chiral
Lagrangian analyses$^{[21]}$. Here, they are found  $\approx 24.5$
and  $\approx .65$ respectively.  Proceeding further, we determine
numerically the KM-matrix for each case seperately. Then, we return
to the neutrino mass matrix and find the mass eigenstates as well as
the diagonalizing  matrix. Then, if $S^{\nu }$ is the matrix which
diagonalizes the effective $3\times 3$ light neutrino sector and
$S_e^L$ the charged lepton mixing matrix, the leptonic mixing matrix
is defined  as follows:
$$V^{lep}=S^{\nu }S_e^{L\dagger }\eqno(39)$$

In the following we present numerical results for some characteristic  values
of the $m_t$ mass. We start running the R.G.E.s from the scale
$M_{GUT}\approx 10^{16}GeV$ (which is known to be the scale where the
standard model gauge couplings meet$^{[22]}$), while the value for
the common gauge coupling at $M_{GUT}$ is taken $g_{GUT}={1\over
{25.1}}$. We will assume that supersymmetry is valid down to the
scale $m_t$ while we run the system with the non-supersymmetric beta
function coefficients bellow $m_t$. First we determine the quark and
charged lepton masses, mixings etc
 which are described in terms of 13 free parameters in the context of the
standard model, only with the eight input parameters ($x,y,z,q, \phi
,a,b,f$) at the GUT scale. Using only two additional inputs which
are the scales of the $\nu ^c \nu ^c $ and $\nu ^c \phi $ entries in
the neutrino mass matrix, we give predictions for the light neutrino
masses and  leptonic mixing angles which can be tested in recent
neutrino experiments. Taking into account all the constraints and
mass relations mentioned above, we present in the following our
results for $ m_t= 130,150$  and $ 160GeV$. We always choose to fix
$a,b$ and $f$ parameters in terms of the charged lepton masses,
hence we give our results only in terms of the set ($x,y,z,\phi$)
and $tan\beta $. Then, $\lambda _t(t_0)$ coupling is also fixed once
$tan\beta $ and $m_t $ are chosen.

For $m_t=130GeV,tan\beta =1.1$ and ${\phi } ={\pi \over 6.5}$, we
obtain the following results:
 $$\eqalign{m_d\approx 6.3MeV, m_s\approx 154MeV, m_b\approx 4.33GeV;\cr
m_u \approx 4.0MeV, m_c\approx 1.27GeV}\eqno(40a)$$
in agreement with the values obtained by the approximation formulae
(28-30) and (34-36).
The Kobayashi-Maskawa matrix elements $\vert (V_{KM})_{ij}\vert $ , are

$$V_{KM}=\left(\matrix{.9754&.2205&.0032\cr
.2202&.9748&.0356\cr
.0108&.0340&.9994\cr}\right)\eqno(40b)$$

For $m_t=150GeV$, we use $tan\beta =2.2$ and ${\phi }={\pi \over
4.5}$.  We get
$$\eqalign{m_d\approx 6.2MeV, m_s\approx 153MeV, m_b\approx
4.25GeV;\cr m_u \approx 4.05MeV, m_c\approx 1.26GeV}\eqno(41a)$$
$$V_{KM}=\left(\matrix{.9752&.2212&.0028\cr
.2109&.9744&.0429\cr
.0117&.0413&.9991\cr}\right)\eqno(41b)$$

Finally, for $m_t=160GeV$, we take $tan\beta \approx 3.3$ and
${\phi } ={\pi \over 4.5}$.Then,
$$\eqalign{m_d\approx 6.2MeV, m_s\approx 152MeV, m_b\approx
4.26GeV;\cr m_u \approx 3.9MeV, m_c\approx 1.26GeV}\eqno(42a)$$
$$V_{KM}=\left(\matrix{.9751&.2219&.0025\cr
.2216&.9741&.0445\cr
.0120&.0430&.9990\cr}\right)\eqno(42b)$$
It is worth noting here, that as the top  mass gets higher the
phase $\phi $ should also become larger in order for the
KM-entries to lie within the experimental limits.
A larger $tan\beta $ is also required.

To obtain the neutrino spectrum and lepton mixing, we must
introduce values for the two additional papameters $M, \Lambda
^{rad} $ of the neutrino mass matrix (3). We assume naturally
$M=<\bar H>\approx 10^{16}GeV$. In order to study the properties of
the neutrino matrix, we let  $\Lambda ^{rad}$ vary in a reasonable
range between $10^{11}$ and $10^{13}$ and fix its value later with
the available neutrino data.

 Next, we parametrize the lepton mixing matrix in a
convenient way,i.e. $$V^{lep}=\left( \matrix{
c_1c_3-s_1s_2s_3e^{i\phi} & s_1c_3+c_1s_2s_3e^{i\phi}& -c_2s_3 \cr
-s_1c_2e^{i\phi} &c_1c_2e^{i\phi}& s_2 \cr
c_1s_3+s_1s_2c_3e^{i\phi}&s_1s_3-c_1s_2c_3e^{i\phi}& c_2c_3  \cr}
\right)\eqno(43)$$

The predictions of the relevant mixing  for the neutrino
oscillations can now be presented in terms of the angles defined in
the parametrization of $V^{lep}$.

In our model described above $V^{lep}$ is fixed by the quark and
charged lepton data. In fact, due to the assumed form of the matrix
$M_{\nu ^c, \phi}$, it only depends on the ratio
$M^{rad}_{33}\over M^{rad}_{11}$ which in our model is equal to
$b\over f$ (see equs (4) and (6a)). The neutrino eigenvalues,
however, cannot be accurately predicted due to the scale quantity
$\Lambda ^{rad}$ which is not specified in our model. Thus they can
be written as
$$m_{\nu _e}\approx 0,
m_{\nu _{\mu}}={\Lambda ^{\mu}\over \Lambda
^{rad}}\times 10^{-2}eV,
m_{\nu _{\tau}}={\Lambda ^{\tau}\over \Lambda
^{rad}}\times 10eV\eqno(44a)$$
For $m_t\approx 130GeV$ we get $\Lambda ^{\mu} \approx .80\times
10^{12}$ and  $\Lambda ^{\tau} \approx 1.85\times 10^{12}$.

Next, we
present the light neutrino  mixing matrices for two
choices of $m_t$, i.e. $m_t=130GeV$ and $m_t=150$ GeV.

For $m_t=130GeV$, we obtain
$$V^{lep}=\left(\matrix{-.9958+7.7\imath \times  10^{-4}&
(-8.5+3.2\imath)\times 10^{-2}
&.00347\cr
(-8.5-3.2\imath)\times 10^{-2}
&.9954+7.7\imath \times 10^{-4}&
-.0307\cr
(-8.4+9.9\imath )\times 10^{-4}&
-.031+.09\imath \times 10^{-3}&-.9995\cr}\right)
\eqno(44b)$$
For $m_t=150 GeV$ we get
$$V^{lep}=\left(\matrix{-.9955+1.4\imath \times  10^{-3}&
(8.4-4.5\imath)\times 10^{-2}
&.0034\cr
(-8.4-4.5\imath)\times 10^{-2}
&-.9947-1.37\imath \times 10^{-3}&
-.0388\cr
(-.13+1.7\imath )\times 10^{-3}&
.039-.098\imath \times 10^{-3}&-.9993\cr}\right)
\eqno(44c)$$

{}From the above, it can be seen that the mixing between all neutrino
species is small. For values of  $\Lambda ^{rad}\approx 10^{12}$ the
obtained neutrino masses are much too small to be detected directly
in present experiments like neutrinoless double beta decay, muon
number violating processes etc. At present, the only place to detect
such small neutrino masses are neutrino oscillation experiments or
astrophysics. We can approximate the oscillation probabilities
relevant to this latter case with a high accuracy in terms of the
$2\times 2$ familiar case, as follows

$$P(\nu _e\leftrightarrow \nu _\mu ) \approx 3.1\times 10^{-2}
sin^2({\pi {L\over l_{12}}})\eqno(45a)$$

 $$ P(\nu _{\tau}\rightarrow \nu _{\mu }) \approx
4.0\times 10^{-3} sin^2({\pi {L\over l_{13}}})\eqno(45b)$$

$$ P(\nu _e\rightarrow \nu _{\tau }) \approx
4.0\times 10^{-5} sin^2({\pi {L\over l_{13}}})\eqno(45c)$$
where L is the source--detector distance and
$$l_{ij}={4\pi E_{\nu}\over |m_i^2-m_j^2|}\eqno(46)$$
Notice that the oscillation length $l_{23}$ does not appear in the
above formulae.
Since however, $m_{\nu _e}\ll m_{\nu _{\mu}}$ and
$m_{\nu _{\mu}}\ll m_{\nu _{\tau}}$, one can in principle constrain
both $m_{\nu _{\mu}}$ and $m_{\nu _{\tau}}$ from such data.  It is
clear from the relations (45a-45c) that our results are
not compatible with large mixing angle experimental limits. Neutrino
oscillations in the medium$^{[23]}$ via the MSW effect$^{[24]}$
provide a solution to the solar neutrino problem.  The GALLEX solar
neutrino data$^{[25]}$
  $$5.0\times 10^{-3}\le sin^22\theta_{ij}\le 1.6\times
10^{-2}\eqno(47)$$
$$0.32\times 10^{-5}(eV)^2\le \delta m_{ij}^2\le 1.2\times
10^{-5}(eV)^2\eqno(48)$$
 can be accomodated in our model. Our result (45a)
is a bit outside the above range but the mass constraint can be
easily satisfied by choosing $\Lambda ^{rad}$ in the range
$$.7\times 10^{12}\le \Lambda ^{rad}\le 7\times 10^{12}$$
Our neutrino masses can also  easily be made to fall into the range
of the Frejus atmospheric neutrinos$^{[26]}$
$$10^{-3}(eV)^2\le \delta m_{ij}^2\le 10^{-2}(eV)^2\eqno(49)$$
but our mixing is much too small. Our results are also consistent
with the data on $\nu _{\mu } \leftrightarrow \nu _{\tau } $
 oscillations$^{[26]}$
$$ sin^22\theta_{\mu \tau}\le 4.\times 10^{-3},\delta m_{\nu _{\mu}
\nu _{\tau}}^2\ge 50(eV)^2\eqno(50)$$
Our results however cannot be made to fall on the $sin^22\theta $
$vs$  $\delta m^2$ of the $BNL$ $\nu _{\mu } \leftrightarrow \nu _e $
oscillation results$^{[27]}$.

Moreover, it is always possible to obtain $m_{\nu _{\tau }}\approx
(few \sim 20)eV$, hence one can obtain simultaneously the
cosmological HOT-dark matter component,
 in agreement with the interpretation of the COBE data$^{[28]}$.
 Indeed
an upper limit on the $\nu _{\tau }$ mass can be obtained from
the formula
$$7.5\times 10^{-2}\le \Omega _{\nu }h^2\le 0.3\eqno(51)$$
Translating this into a constraint on $m_{\nu {_\tau}}$, arising
from the relation $m_{\nu_{\tau }}\approx \Omega _{\nu }h^291.5eV$
where $h=.5\sim 1.0$ is the Hubble parameter, one gets the range
     $$6.8\le m_{\nu _{\tau }}\le 27eV\eqno(52)$$
 which can be easily achieved with the above range of $\Lambda
^{rad}$.

In conclusion, we have proposed a structure of the fermion mass
matrices in the flipped SU(5) model. By allowing the Yukawa
couplings to evolve from the GUT scale down to $m_W$, using
only 8 input parameters at the GUT scale,
we can fix
all the 13
measurable parameters (masses and mixings angles),at
$m_W$. Furthermore, with the above information our model allows us
to make definite predictions for the neutrino masses and the
leptonic ``Kobayashi Maskawa'' matrix.  In particular, we have found
that the  generalized  see-saw mechanism which occurs naturally in
this model can provide a solution to the solar neutrino problem via
the MSW mechanism. Moreover, a sufficiently large  $\nu _{\tau }$
mass is always possible in this model in order to contribute as a
hot dark matter component,  as indicated by the recent COBE data.

{\it Note added}. After the completion of this work we received a
copy of the paper of ref[29], where it is shown that the
generalized see-saw mechanism in this model can also account for
the baryon asymmetry of the Universe. Indeed, in our model we
also have at least one singlet neutrino state with mass of order
$10^{11}GeV$.

\vskip 1.0cm
{\it {\bf Acknowledgements }}
\vskip 0.2cm
{\it We would like to acknowledge
partial support by the EC-grant SCI-0221-C(TT)}
 \vfill

 \eject
\par
 \vskip -1.0cm
\centerline{\underbar{\sl REFERENCES}}
\openup2pt{$$\vbox{\settabs\+\indent&[12]\quad&S.M. Bilenky, A.
Masiero and S.T. Petcov, Phys.Lett.B  66(1991)2444;\qquad\cr
\+[1]&K.S. Hirata et al Phys. Rev Lett. 65(90) 1297;\cr
\+ &K.S. Hirata et al, Phys. Rev. D44 (91) 2141 ;\cr
\+ &SAGE collaboration A.I.Abazov et al, Phys.Rev.Lett.67(91)
3332;\cr
\+[2] &K.S. Hirata, et al, Kamiokande-II Collaboration
(Phys.Lett.B) ;\cr
 \+[3]& R Becker-Szendy et al Boston Univ. preprint
BUHEP-91-2491992)\cr
\+[4]&G.F. Smooth et al,Astrophys.J. Lett. (to appear).\cr
\+[5]&JR.K. Schaefer and Q.Shafi Ic/92/118 (BA-92-45);\cr
\+[6]&T.Yanagida, Prog. Th. Phys. B 135 (1978) 66 ;\cr
\+& M. Gell Mann P. Ramond and Slansky, ;\cr
\+&in Supergravity,ed P.van Niewenhuizen and D. Freedman ;\cr
\+&(North Holland 1979 p.315);\cr
\+&R. Mohapatra and G. Senjanovic, Phys.Rev.Lett.44(1980)912.\cr
\+[7]&S. Dimopoulos, L. J. Hall and S. Raby,\cr
\+& Phys. Rev. Lett.68(92) 1984; Phys. Rev D 45 (92) 4192;\cr
\+& V. Barger et al Phys. Rev. Lett 68 (1992)3394.\cr
\+[8]& G. F. Giudice, UTTG-5-92, March 1992, Texas Univ. Preprint;\cr
\+&R.G. Roberts and G.G. Ross, Oxford Preprint (92);\cr
\+& K.S.Babu and Q. Shafi, Univ. of Delaware preprint.\cr
\+[9]&S.Dimopoulos, L.J. Hall and S. Raby LBL-32484/92,August.\cr
\+[10]&H. Dreiner, G.K. Leontaris and N.D. Tracas, Oxford;\cr
\+&preprint,September 1992;IOA-281/92\cr
\+[11]&I. Antoniadis et al, Phys. Lett. B 194 (1987) 231;\cr
\+[12]&I. Antoniadis and G.K. Leontaris Phys. Lett. B 216(1989)
333;\cr
\+[13]& G.K. Leontaris Phys. Lett. B 207(1988)447;\cr
\+& G.K. Leontaris and D.V. Nanopoulos Phys. Lett. B 212(1988)327;\cr
\+&G.K. Leontaris and C.E. Vayonakis, Phys. Lett. B206(1988)271;\cr
\+&S. Abel, Phys.Lett.B234(1990)113;\cr
\+&I. Antoniadis, J.Rizos and K.Tamvakis Phys.Lett.B279(1992)281;\cr
\+[14]&G.K. Leontaris and J.D.Vergados Phys. Lett. B258(1991)111;\cr
\+&E. Papageorgiu and S. Ranfone Phys.Lett.B282(1992)89;\cr
\+[15]&J. Ellis, J.L. Lopez and D.V. Nanopoulos
Phys.Lett.B(1992);\cr
\+[16]&S.Kalara, J.L. Lopez and D.V. Nanopoulos,
Phys.Lett.B245(1990)421;\cr
 \+[17]&J.W.F. Valle,''Gauge theories and
the physics of
 neutrino mass;\cr
\+&Progress in Part. and Nucl. Phys. Vol.26,1991.\cr
 \+[18]&E.Witten, Phys.Lett.B91(1980)81;\cr
}$$}

\vfill
\eject
\par
\openup2pt{$$\vbox{\settabs\+\indent&[12]\quad&S.M. Bilenky, A.
Masiero and S.T. Petcov, Phys.Lett.B  66(1991)2444;\qquad\cr

 \+[19]&R. Barbieri et
al., Phys.Lett.B155(1982)212;\cr
\+&M.B.Einhorn and
D.R.T.Jones, Nucl.Phys.B196(1982)457;\cr
 \+&J.P.Derendinger and C.A.
Savoy, Nucl.Phys.B237(1984)307;\cr \+&B.Gato et al.,
Nucl.Phys.B253(1985)285;\cr  \+&N.K. Falck, Zeit. Phys. C 30(1986)
247.\cr
 \+[20]&J.Gasser and H. Leutwyler, Phys.Rep.87(1982)77\cr
\+[21]&D. Kaplan and A. Manohar, Phys.Rev.Lett.56(1986)2004;\cr
\+&H. Leutwyler, Nucl.Phys.B337(1990)108.\cr
\+[22]&J. Ellis, S. Kelley and D.V. Nanopoulos, Phys. Lett. B 249
(1990)441;\cr
\+& P. Langacker and M. Luo, Phys. Rev. D 441 (1991) 817; \cr
 \+&U.Amaldi, W. de Boer, and H. F\"urstenau, Phys. Lett. B 260
(1991) 447;\cr
\+ &R.G. Roberts and G.G. Ross, Nucl. Phys. B 377 (1992) 571.\cr
\+[23]&P.I. Krastev and S.T. Petcov,CERN-TH 6539/92;\cr
\+&S.M. Bilenky, Neutrino mixing DFTT67/92.\cr
\+[24]&L. Wolfenstein, Phys.Rev.D17(1978)2369;\cr
\+&S.P. Mikheyev and A.Yu. Smirnov, Yar.Fiz.42(1985)1441.\cr
\+[25]&P. Anselmann et al. Phys.Lett.285(1992)376\cr
\+[26]&Ch. Berger et al Phys. Lett.B245(1990)305\cr
\+[27]&L. Borodovsky et al, Phys.Rev.Lett.68(1992)247\cr
\+[28]&F.L. Wright et al. Ap.J.Lett.396(1992)L13\cr
\+[29]&J. Ellis, D.V. Nanopoulos and K.A. Olive,\cr
\+&``Flipped heavy neutrinos: from the solar neutrino problem \cr
 \+&to baryogenesis'' CERN-TH6271/92,ACT-22/92,UMN-TH1117/92\cr }$$}
\vskip 1.5in
\vfill
\eject
\centerline{{\bf TABLE 1.}}

$$\vbox{\settabs 3 \columns
\+${\bf m_t=130GeV}$&{}&{}\cr
\+${\bf \Lambda _{rad}}$ & ${\bf m_{\mu}}$ & ${\bf m_{\tau }}$\cr
\+$0.50\times 10^{12}$ &$1.61\times 10^{-2}$ &36.9eV\cr
\+$0.75\times 10^{12}$ &$1.08\times 10^{-2}$ &24.6eV\cr
\+$1.00\times 10^{12}$ &$0.80\times 10^{-2}$ &18.5eV\cr
\+$3.00\times 10^{12}$ &$0.27\times 10^{-2}$ &{}6.2eV\cr
\+$5.00\times 10^{12}$ &$0.16\times 10^{-2}$ &{}3.7eV\cr
\+$7.00\times 10^{12}$ &$0.12\times 10^{-2}$ &{}2.6eV\cr}$$
$$\vbox{\settabs 3 \columns
\+${\bf m_t=150GeV}$&{}&{}\cr
\+${\bf V^{lep}_{e\mu }}$&${\bf V^{lep}_{\mu \tau}}$&
${\bf V^{lep}_{\tau e}}$\cr
\+$(8.38-\imath 4.46)\times 10^{-2}$&
$3.88\times 10^{-2}$&
$(-.125+\imath 1.74)\times 10^{-3}$\cr
\+${\bf \Lambda _{rad}}$ & ${\bf m_{\mu}}$ & ${\bf m_{\tau }}$\cr
\+$0.50\times 10^{12}$ &$1.68\times 10^{-2}$ &59.7eV\cr
\+$0.75\times 10^{12}$ &$1.12\times 10^{-2}$ &39.8eV\cr
\+$2.00\times 10^{12}$ &$0.42\times 10^{-2}$ &14.9eV\cr
\+$5.00\times 10^{12}$ &$0.17\times 10^{-2}$ &{}5.9eV\cr
\+$7.00\times 10^{12}$ &$0.12\times 10^{-2}$ &{}4.3eV\cr}$$


 \eject
\bye